\documentclass{KapProc} 

\let\footnote\savefootnote
\let\footnotetext\savefootnotetext 
\let\footnoterule\savefootnoterule 

\kluwerbib

\startauthorindex

\usepackage{graphicx} 

\begin{document}



\articletitle[]{Morphological Properties of PPNs:
Mid-IR and HST Imaging Surveys}

\chaptitlerunninghead{Morphological Properties of PPNs:
mid-IR and HST}

\author{Toshiya Ueta and Margaret Meixner}
\affil{University of Illinois at Urbana-Champaign}
\email{ueta@astro.uiuc.edu, meixner@astro.uiuc.edu}

\begin{abstract}
We will review our mid-infrared and HST imaging surveys
of the circumstellar dust shells of proto-planetary nebulae.
While optical imaging indirectly probes the dust distribution 
via dust-scattered starlight, mid-IR imaging directly
maps the distribution of warm dust grains.
Both imaging surveys revealed preferencially axisymmetric 
nature of PPN dust shells, suggesting that 
axisymmetry in planetary nebulae sets in by the
end of the asymptotic giant branch phase, most likely
by axisymmetric superwind mass loss.
Moreover, both surveys yielded two morphological
classes which have one-to-one correspondence between
the two surveys, indicating that the optical depth of
circumstellar dust shells plays an equally important
role as the inclination angle in determining the 
morphology of the PPN shells.
\end{abstract}


\section{Introduction}

While circumstellar dust shells (CDSs) of 
asymptotic giant branch (AGB) stars display
high spherical symmetry (e.g. Habing \& Blommaert 1993),
planetary nebulae (PNs) are renowned for their
spectacular axisymmetry (e.g. Zuckerman \& Aller 1986).
To understand {\sl when} and {\sl how} such a
drastic morphological conversion takes place,
the proto-planetary nebula (PPN) phase, a brief 
transitionary phase between the AGB and PN phases (e.g. Kwok 1993)
has been extensively investigated.
In fact, PPNs are the most ideal space laboratories
for such studies:
AGB mass loss histories are imprinted on the PPN 
shells in the most pristine form 
because the PPN central stars are 
not hot enough to generate disturbing 
fast winds and/or ionizing UV photons
as PN central stars.

Therefore, by conducting imaging surveys on a 
large number of PPN CDSs, 
we can pinpoint the epoch of morphological 
transformation and obtain insights about
yet unknown mechanisms for the AGB mass loss.
To achieve our goals, we employed 
mid-infrared (IR) and optical wavelength ranges,
both of which have strengths and weaknesses
that would complement one another.  
Diffraction-limited mid-IR images give
marginal resolution ($1^{\prime\prime}$) but allow one to 
directly probe dust mass distribution through 
thermal dust emission.
On the other hand, 
optical images are of high resolution ($0^{\prime\prime}.1$),
while they only allow an indirect mass probing through
dust-scattered starlight escaping from the dust shells.

\section{Mid-IR Imaging Survey}

The aim of the mid-IR survey is to map out 
distribution of warm dust grains at the inner edge
of the PPN CDSs. 
So far we have observed about 70 PPN candidates with
8 to 25 micron filters.
We have been using IRTF and UKIRT 
($\sim 1^{\prime\prime}$ resolution with $\sim 3$m apertures) 
and MMT
(sub-arcsecond resolution with 6.5m aperture).
Out of 72 objects, we have only resolved 22 object
due to marginal resolution.
However, 16 of these 22 showed clear elongation, and
moreover, elongations were morphologically characterized
into two groups: toroidal and core/elliptical types
(Meixner et al.\ 1999).
	 
\begin{figure}[h]
\includegraphics[width=13.6pc]
{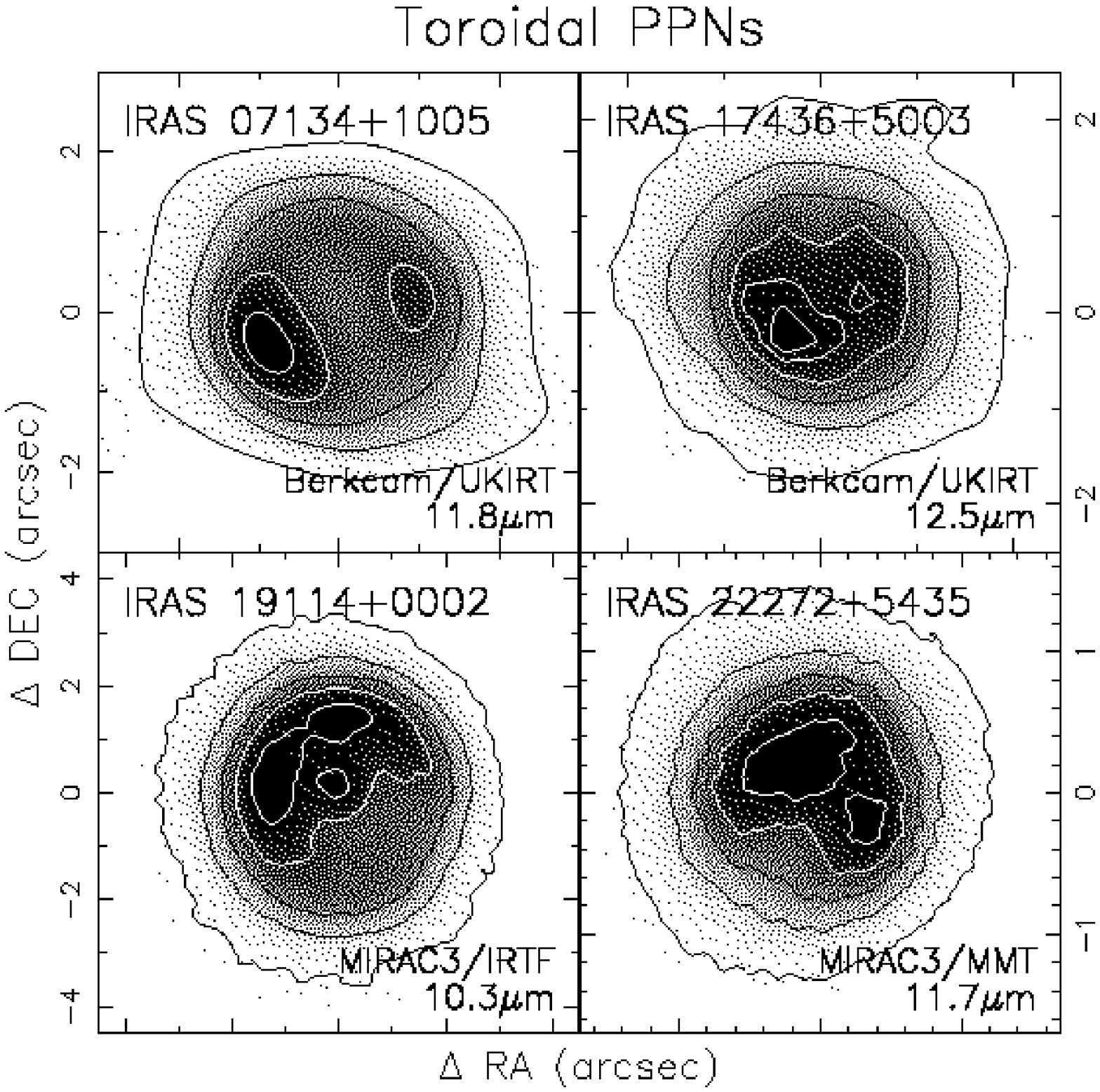}
\includegraphics[width=13.6pc]
{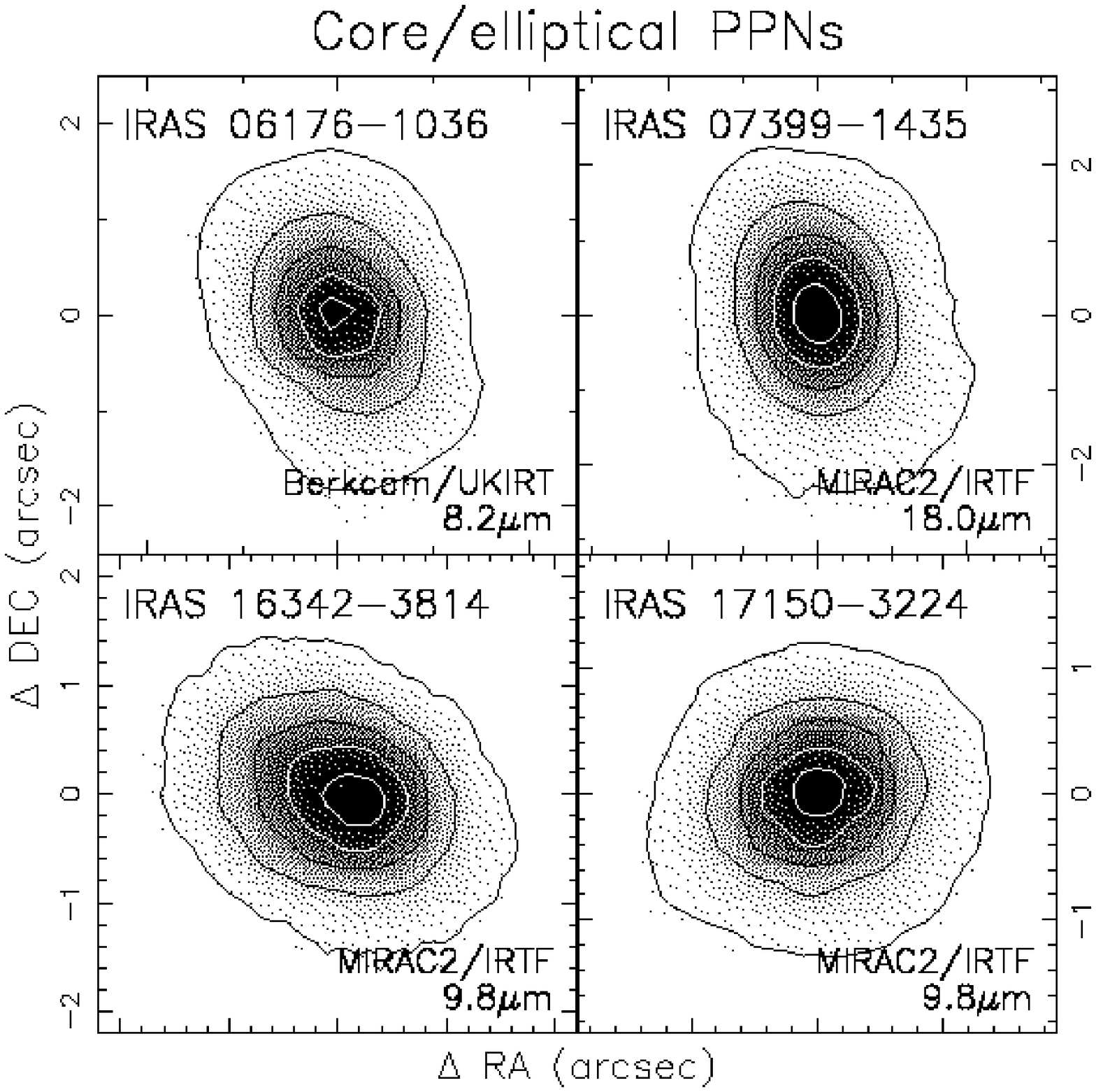}
    \caption{Two types of mid-IR morphologies:
toroidal (left) and core/ellitical (right) PPNs.
Indicated are object names, instruments, wavelengths, 
and sizes.  
The size of the emission core with respect to the entire 
nebula is the main difference.}
\end{figure}

The toroidal PPN morphology is characterized
by a large emission core which shows some evidence 
of a dust torus.
The resolved peaks in the emission core are interpreted 
as limb-brightened edges of an optically thin dust torus.
i.e., the inner structure is revealed through optically 
thin dust shells.
The morphology of the core/elliptical PPNs is characterized
by a relatively compact emission core which is surrounded by
a larger emission plateau.
This type of PPNs seem to keep their inner structure
hidden beneath optically thick dust shells.

\section{HST Imaging Survey}

The goal of the optical imaging survey was to 
detect faint circumstellar nebulosities
caused by dust-scattered starlight.
We observed 27 PPN candidates using HST/WFPC2 during 
cycle 6 with typical resolution of $0^{\prime\prime}.1$.
21 out of 27 objects showed nebulosities,
and all of which showed two kinds of axisymmetry
(Ueta et al.\ 2000):
the first group being star-obvious low-level elongated
nebulae (SOLE) and the second group being 
dust-prominent longitudinally extended nebulae (DUPLEX).

\begin{figure}[h]
\includegraphics[width=13.6pc]
{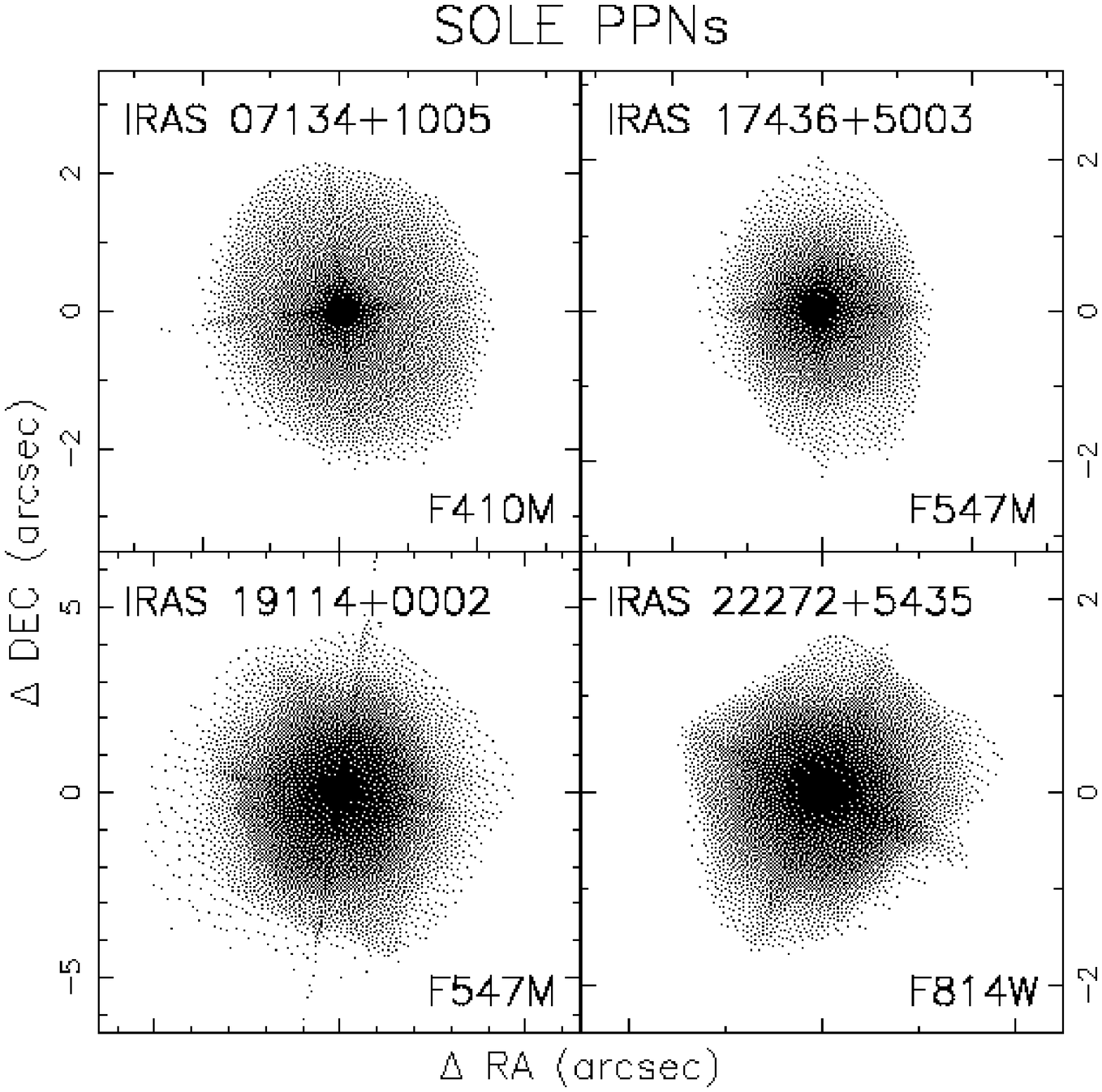}
\includegraphics[width=13.6pc]
{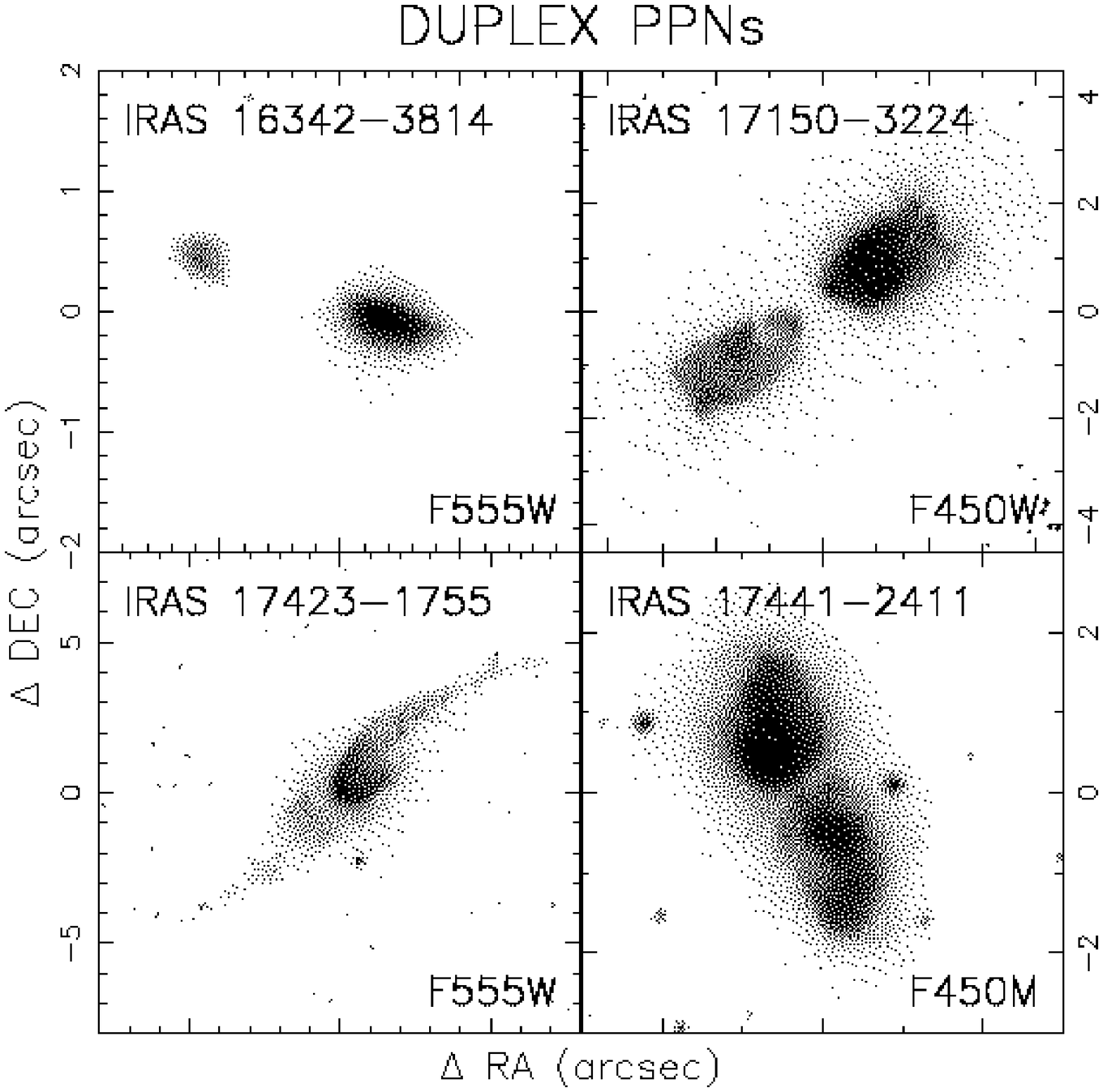}
\caption{Two types of optical morphologies:
SOLE (left) and DUPLEX (right) PPNs.
Indicated are object names, WFPC2 filters, 
and sizes.  
The main difference is the visibility of the central star
due to dust obscuration, which manifests itself in the 
nebula shapes.}
\end{figure}

SOLE PPNs have very faint and smooth, ellitpically
elongated nebulosities with the central stars 
prominently visible at the very center of the nebulae. 
The morphology of DUPLEX PPNs is characterized
by their bipolar nebulosities.
Depending on the inclination angle of the object,
you may or may not see the central star but
you will always recognize some evidence for
a dust lane between lobes.

\section{Combined Results}

Both surveys showed that PPN dust shells have
a striking preference towards axisymmetry.
This is consistent with previous studies on 
the axisymmetry in PPNs (e.g., Trammel et al.\ 1994;
Hrivnak et al.\ 1999).
It is, therefore, most likely that
the epoch of morphological transformation
should start before the end of the AGB phase.
If we restrain ourselves from invoking any exotic 
mechanisms,
a standard AGB evolution scenario (e.g., Iben 1995
for a review) suggests one event which might trigger 
the tranformation, the superwind.  
The bulk of the PPN dust shell is considered to
be ejected during this brief but violent phase of 
mass loss at the end of the AGB phase, 
and superwind may be violent enough to cause 
such a drastic morphological change.

\vspace{0.3pc}
\begin{minipage}{13.75pc}
\includegraphics[width=13pc]
{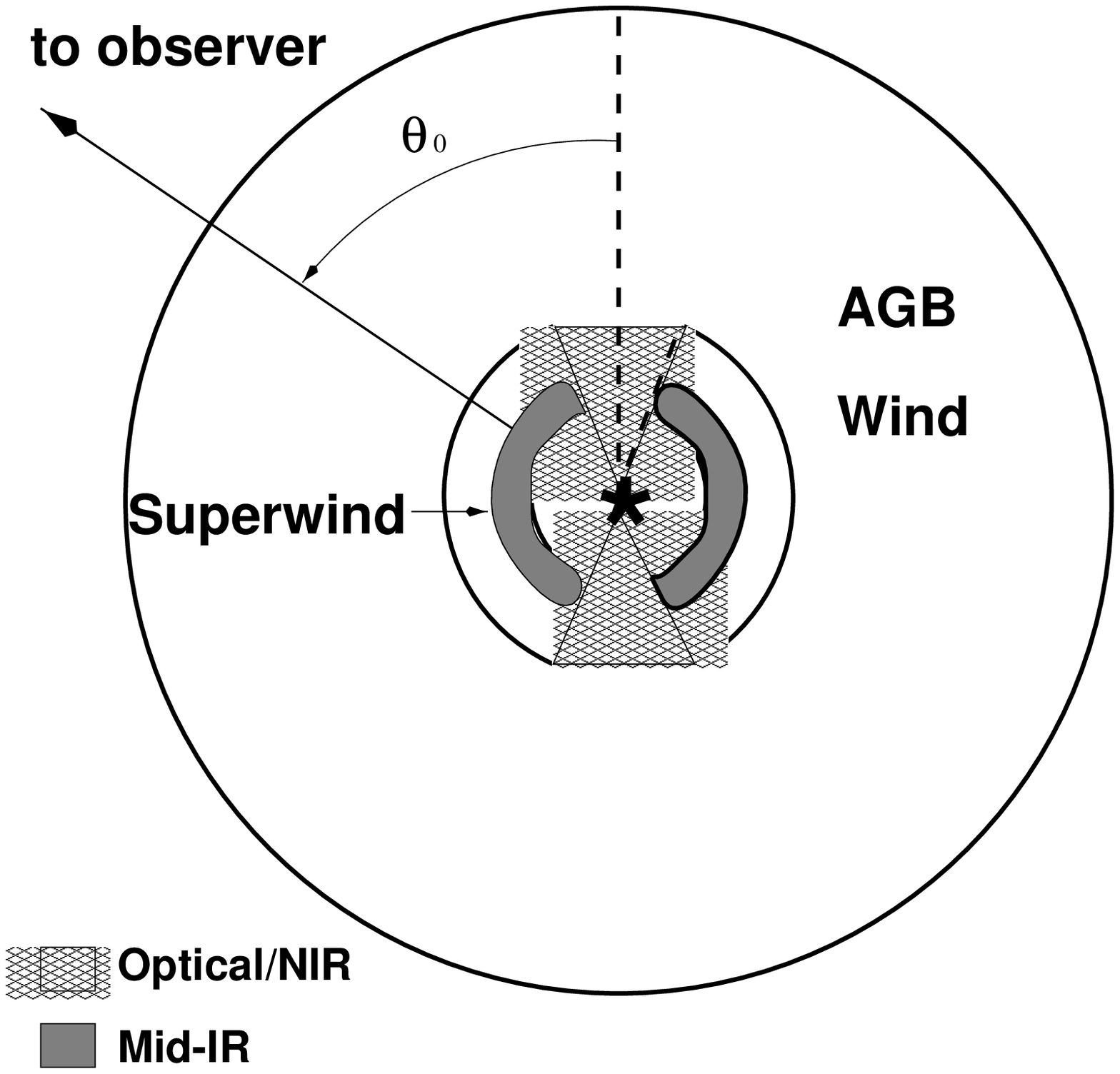} 
\end{minipage}
\begin{minipage}{13.75pc}
~~Thus, we propose a zeroth-order model for the structure 
of the PPN dust shells, shown on the left.
In this model, an axisymmetric superwind shell is
embedded in an otherwise spherical AGB wind shell.
Mid-IR thermal emission comes from the inner edge
of the superwind shell, which manifests itself as 
a dust torus we have observed (Fig. 1), 
while optical bipolar lobes are evidently
dust-scattered starlight escaping
the dust torus through these bicone openings.
\end{minipage}
\vspace{0.01pc}

Now, how does this model explain the observed 
dual morphologies in these surveys?
Interestingly, there is a one-to-one correspondence 
between two dual morphologies found in both surveys
(Fig. 3):
toroidal mid-IR shells are always associated with
SOLE type optical nebula, and core/elliptical mid-IR
shells are always associated with DUPLEX type 
optical nebulosities.

\begin{figure}[t]
\includegraphics[width=14.3pc]
{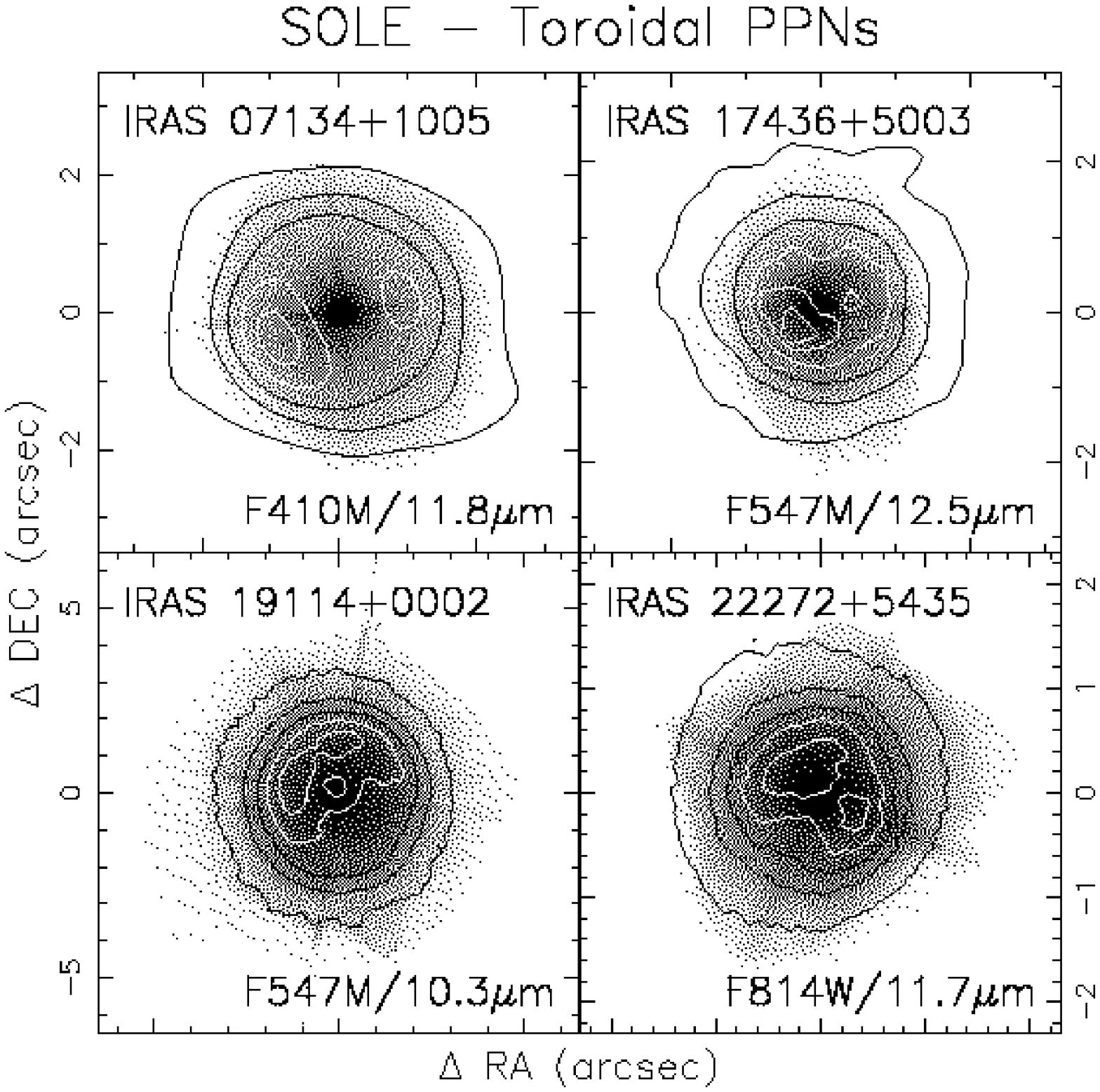}
\includegraphics[width=14.3pc]
{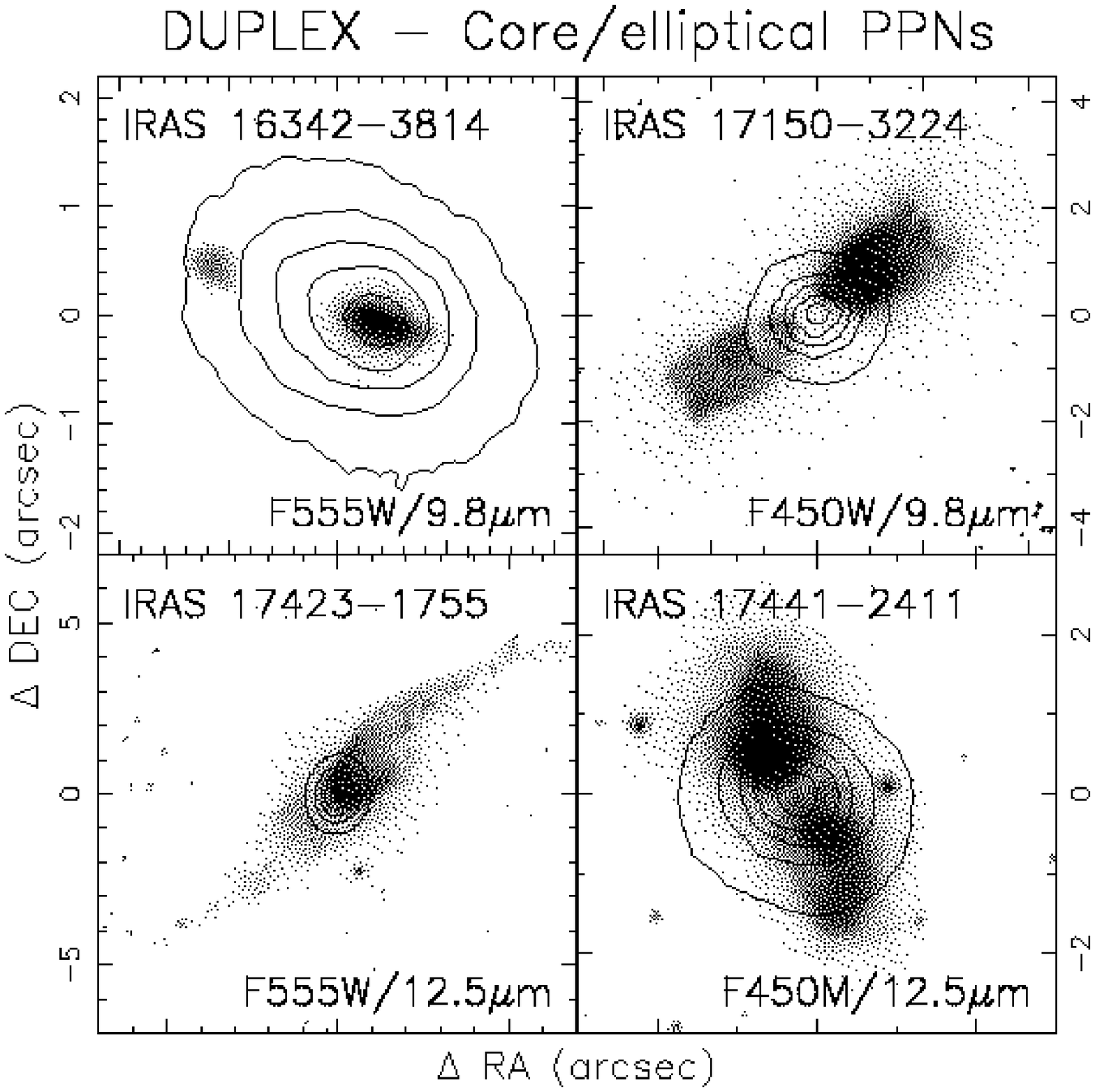}
\caption{Two PPN types?: 
grayscale HST images overlaid with Mid-IR contours
showing a one-to-one correspondence.}
\end{figure}

\begin{figure}[b]
\includegraphics[width=27.25pc]
{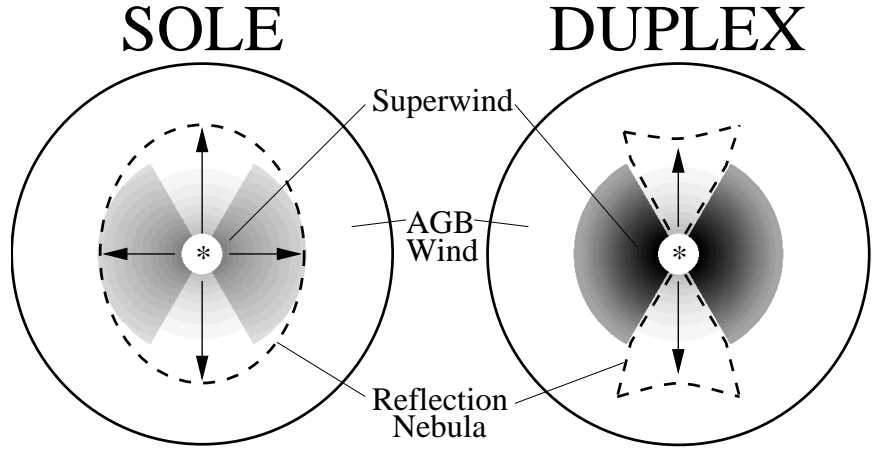}
\caption{Optical depth interpretation of the 
mid-IR/optical dual morphologies of the two PPN types.
SOLE-Toroidal (left) is a manifestation of 
an optically thin PPN dust shell
while DUPLEX-Core/elliptical (right) is
a result of an optically thick PPN dust shell.}
\end{figure}

While the extent of both mid-IR and optical nebulae is
comparable in SOLE-Toroidal PPNs, in
DUPLEX-Core/elliptical PPNs, optical nebulae
can reach far beyond the mid-IR emitting regions, 
which roughly cover the core of the dust torus.
The optical morphologies have been explained
by the inclination angle effect, 
in which an intrinsically bipolar nebula can be
viewed as a bipolar nebula near edge-on or 
an ellipitically elongated nebula near pole-on
depending on the the inclination angle of the object.
This explanation fails to explain the mid-IR
morphology of SOLE-Toroidal PPNs.
Within the framework of the inclination angle interpretation, 
SOLE-Toroidal PPNs must be oriented nearly pole-on.
However, their mid-IR morphology clearly indicates 
the presence of rather edge-on dust toroids
and can not be satisfactorily explained 
by the inclination angle effect alone.

The two-layered PPN shell model we proposed can
explain both mid-IR and optical morphologies
if we consider the PPN dust shells that can have
a range of optical depth (Fig. 4).
In this optical depth effect interpretation,
the optical depth of the SOLE-Toroidal PPN dust 
shells are low enough that starlight can scatter 
into all directions making an elliptical looking 
nebulosities while mid-IR images showing 
two emission peaks.
On the other hand, in DUPLEX-Core/relliptical PPNs, 
the optical depth is very high and starlight can
escape into only the bicone openings of the dust 
shell making bipolar 
nebulosities while mid-IR emission showing
only one, compact emission core.

Although the overall morphology of PPN dust shells
is still dependent on the inclination angle of the 
shells, the results from our surveys evidently demonstrated
that the optical depth effect plays an equally important 
role in defeining the morphology of the PPN dust shells.

\section{Future Prospects}

As we have seen, well-resolved mid-IR images are very
effective means to directly probe the structure of 
dust distribution.  
With large aperture telescopes coming online, we are now 
able to obtain sub-arcsecond resolution mid-IR images,
and detailed spatial information at the innermost
regions of PPN CDSs will permit refined understanding
of the AGB mass loss.
Also, spatially detailed data will help constraining 
input geometrical parameters for radiative transfer 
calculations done in a fully two-dimensional grid.
These model calculations are important because
they allow us to quantify the axisymmetric nature of
PPN CDSs by, for example, the pole-to-equator density 
ratio.  
We will also be able to constrain the inclination 
angle from such model calculations, clarifying ambiguities
between effects of the inclination angle and optical depth
of the PPN dust shells.


\begin{acknowledgments}
These surveys have been done in collaboration with 
W. F. Hoffmann, P. Hinz (Steward Obs./Univ. of Arizona),
A. Dayal (IPAC/JPL),
J. L. Hora, G. Fazio (Harvard/CfA),
L. K. Deutsch (Boston Univ.),
B. J. Hrivnak (Valparaiso Univ.),
C. J. Skinner (deceased), and
M. Bobrowsky (Challenger Center for Space Science Education).
\end{acknowledgments}


\end{document}